# Manipulation of the magnetic order parameter and the metal-insulator-transition of a manganite thin film with applied elastic stress

Surendra Singh,[1] M. R. Fitzsimmons,[2] T. Lookman,[2] H. Jeen[3,4] and A. Biswas[3]

[1] Solid State Physics Division, Bhabha Atomic Research Center, Mumbai 400085, India

[2] Los Alamos National Laboratory, Los Alamos, NM 87545, USA

[3] Department of Physics, University of Florida, Gainesville, FL 32611, USA

[4] Department of Physics, Pusan National University, Busan 609-735, Korea

**Abstract:** We measured the temperature dependence of the saturation magnetization ($M_s$) of a $(La_{1-x}Pr_x)_{1-y}Ca_yMnO_3$ (x ~ 0.60, y ~ 0.33) film as a function of applied bending stress. Stress producing a compressive strain of -0.01% along the magnetic easy axis increased the Curie temperature by ~6 K and the metal-insulator-transition by ~4 K. Regardless of whether or not stress is applied to the film, magnetic ordering occurs at temperatures significantly higher than the metal-insulator-transition temperature. The magnetization of the sample at the temperature of the metal-insulator-transition is approximately the site percolation threshold for a two-dimensional spin lattice.

Hole doped manganites are strongly correlated oxides that show a large variety of magnetic and electronic phases due to competing interactions between the orbital, charge, lattice and spin degrees of freedom.[1,2,3] The competing interactions lead to complex behavior such as colossal magnetoresistance (CMR) often accompanied with metal-to-insulator transitions (MIT).[4,5,6,7] Competition between the interactions can be influenced by electric fields,[8,9] magnetic field,[6] light,[10] stress/strain,[3,11,12,13,14,15] disorder.[15] Among these parameters, stress/strain is ubiquitous in thin films and devices, and can significantly affect the properties of manganite thin films.[3,11,12,13,14,16,17,18,19,20] Millis *et al.*,[18] proposed an analytical model to describe the effects of biaxial strain on the transport properties of CMR manganites. For example, one prediction is that 1% biaxial strain can cause a 10% shift of the Curie temperature ($T_c$).

Although the complex phase diagram of manganites has been explored in light of strong coupling between the electronic and elastic degrees of freedom, crucial features remain very divisive. Notable examples are the cluster-glass-like magnetism and a possible cooperative transition from a *strain-liquid* to a *strain-glass* phase at low temperature.[21,22] Closely related to these phenomena is the nonstandard percolative nature of the MIT[1,2,3,4,5,6,7,23] which remains a long-standing controversial problem.

Recently, we observed a strong influence of applied bending stress on the saturation magnetization, $M_s$, and the MIT of a manganite film.[24] Namely, compressive stress increases $M_s$ and the MIT. However the exclusive influence of applied elastic stress on $T_c$ of a manganite thin film has not been quantitatively reported. Here, we report the effect of applied elastic bending stress on the magnetic ordering temperature of a single crystalline $(La_{1-x}Pr_x)_{1-y}Ca_yMnO_3$ (LPCMO) film. Significant increases of both the MIT and $T_c$ with small (0.01%) compressive strain mean that changes of strain alone are sufficient to affect the electronic properties and magnetic ordering in LPCMO thin films. However, regardless of whether or not stress was applied to the sample, magnetic ordering occurs at significantly higher temperatures than the MIT, well within the insulating phase.

A 20-nm-thick LPCMO film was epitaxially grown on a 1cm by 1 cm (110) NdGaO$_3$ (NGO) substrate in the step-flow-growth-mode using pulsed KrF laser (248 nm) deposition (PLD).[25] During growth, the substrate temperature was 1053 K, O$_2$ partial pressure was 17.33 Pa, laser fluence was 0.5 J/cm$^2$, and the repetition rate of the pulsed laser was 5 Hz. The thickness of the substrate was 0.25 mm. The lattice mismatch between bulk LPCMO and NGO is about 0.26% and 0.49% along [1$\bar{1}$0] and [001] NGO, respectively.[25] Since we are examining one sample as a function of *applied bending stress*, the so-called epi-strain resulting from lattice mismatch is not affected by our experiment. For the purposes of this paper, strain means the strain resulting from the bending stress applied to the sample.

Bending stress was applied using a four point mechanical jig, which applied stress uniformly over the lateral dimensions of a large sample.[24] We simultaneously measured the polarized neutron reflectivity[26,27] (PNR), $\mathscr{R}$, of the sample (Fig. 1) and resistance, R, along the 1-cm-length of the sample [Fig. 2(a)] as functions of temperature, applied magnetic field and applied stress.[24] Bending stress was applied parallel to the magnetic easy axis of the sample (i.e., parallel to [1$\bar{1}$0] NGO).[24,28] The strain of the film was obtained by measuring the radius of curvature of the film using a laser.[24] In contrast to V$_2$O$_3$, which is metallic above and insulating below its MIT,[29] LPCMO is insulating above and metallic below its MIT. The maximum of dR/dT yields the metal-to-insulator transition temperature $T_{MI}$ = 101.4 K without applied stress and 105.4 K with applied compressive stress. $T_{MI}$ represents the temperature at which percolation of the metallic phase is detected in a film with effectively infinite lateral dimensions. Small compressive bending stress (i.e. $\varepsilon \sim -0.01\%$) induces a positive shift of $T_{MI}$ by $\sim 4$ K [Fig. 2(a)]. Later we show $T_c$ is ~18 K greater than the MIT. Because the transport measurements were made during the neutron experiment, the difference between the $T_{MI}$ and $T_c$ cannot be due to errors in thermometry.

In order to probe the depth dependent structure and magnetization of the film, we carried out specular X-ray reflectivity (XRR) and PNR measurements.[26,27] The specular reflectivity, $\mathcal{R}$, of the sample was measured as a function of wave vector transfer, $Q = 4\pi \sin\theta/\lambda$ (where, $\theta$ is angle of incidence and $\lambda$ is the x-ray or neutron wavelength). The reflectivity is qualitatively related to the Fourier transform of the scattering length density (SLD) depth profile $\rho(z)$, averaged over the whole sample area. In case of PNR, $\rho(z)$ consists of nuclear and magnetic SLDs such that $\rho^{\pm}(z) = \rho_n(z) \pm CM_s(z)$, where $C = 2.91\times10^{-9}$ Å$^{-2}$(kA/m)$^{-1}$ and $M_s(z)$ is the saturation magnetization (in kA/m) depth profile.[26] The +(−) sign denotes neutron beam polarization along (opposite to) the applied field. $\rho_n(z)$ and $M_s(z)$ can be inferred from $\mathcal{R}^{\pm}(Q)$ often with nm+ resolution. The reflectivity data were normalized to the Fresnel reflectivity[26] ($\mathcal{R}_F = \frac{16\pi^2}{Q^4}$).

The XRR (Fig. S1 in supplementary information[30]) guides modeling of the film's chemical (or nuclear) structure, *e.g.,* film thickness, roughness, *etc*. Previously, scanning transmission electron energy-loss spectroscopy (EELS) microscopy[24,28] found the composition of an identically prepared film to be (La$_{1-x}$Pr$_x$)$_{1-y}$Ca$_y$MnO$_3$(x ~ 0.55, y ~ 0.23) averaged over the entire thickness of the sample. The average composition, however, does not adequately represent the significant variation of the chemical depth profile. The EELS measurement and XRR measurements of the present sample suggest three chemically distinct regions (surface, bulk-film and film-substrate regions) which are manifested in the variation of the x-ray scattering length density depth profile (and the nuclear scattering depth profile used in the neutron analysis, discussed later).[24,28]

The PNR measurements (Fig. 1) were carried out using the Asterix spectrometer at LANSCE.[26] A field of 6 kOe was applied along easy axis during and after cooling the sample at a rate of 0.40±0.05 K /min. Figs. 1(a) and (b) show $\mathcal{R}^{\pm}(Q)$ in the absence of compressive strain at 170 K (in the non-magnetic phase) and 20 K, respectively. At 170 K, the reflectivities measured with and without applied stress were statistically the same. The inset of Fig. 1(b) shows $\mathcal{R}^{\pm}(Q)$ at 20 K for the compressive strain condition ($\varepsilon \sim -0.01\%$).

Reflectivity data were analyzed using the dynamical method of Parratt.[26,31] The nuclear SLD shown as the solid (black) curve in Fig 1(c) is an optimal fit to $\mathcal{R}^{\pm}(Q)$ [solid (black) curves in Figs 1(a)] at 170 K. The three regions with different nuclear (chemical) SLD's are represented as I (surface), II (bulk-film) and III (film-substrate interface) in Fig. 1(c). Next, the nuclear SLD was fixed and then *only* the three values of $M_s$ corresponding to Regions I, II and III were optimized to the PNR data taken at 20 K. Fig. 1(d) shows $M_s(z)$, as the solid (black, $\varepsilon = 0$) and dash (blue, $\varepsilon = -0.01\%$) depth profiles yielding the curves in Fig. 1(b). The fits assumed the same roughness for the magnetic and nuclear interfaces. Because the fit was already very satisfactory, further adjustable parameters, *e.g.* different values for magnetic and nuclear roughness, are not warranted. Regardless of variation of the chemical composition across the film's thickness, $M_s$ is larger for the strained film compared to the unstrained film. This result is consistent with the previous study of an identically prepared sample.[24]

We next investigated the *temperature dependence* of the $M_s$ for conditions without and with compressive strain. PNR data were collected for a range of $Q$ extending to 0.032 Å$^{-1}$ in order to maximize the number (thirty-two) of temperature measurements (see Fig. S2 in supplementary information[30]). Fitting only $M_s$, we obtained the results shown in Fig. 2 for Region II.

The resistance measurements of the sample for the two states of strain are superimposed on the $M_s(T)$ results. At 120 K the sample is insulating and yet retains ~60% of its 20 K saturation magnetization, i.e., $M_s(T_{MI})/M_s(20\ K) \sim 0.6$. These data lead us to conclude that magnetic ordering occurs in the film at temperatures much higher than the MIT (for a sample with effectively infinite lateral dimensions) whether or not stress is applied to the film. The value ~ 0.6 is consistent with the site percolation thresholds for *two-dimensional* lattices which range from 0.5 (triangular lattice),[32] 0.59 (square lattice) [33] to 0.7 (honeycomb lattice),[34,35] and is inconsistent with site percolation thresholds for three-dimensional lattices which are considerably smaller. We suggest the onset of the metal-insulator-transition may be related to percolation of the magnetic phase if the magnetic phase is also metallic. (The bond percolation threshold, which should apply to transport of current, varies between 0.35 and 0.65 for various two-dimensional lattices.[35])

$M_s(T)$ near the magnetic ordering temperature is shown in Fig. 2(b). To estimate values of $T_c$, we fitted $M_s$ to a line in the temperature region between 120 K and 130 K. Extrapolating the line to $M_s = 0$ yields an estimate of $T_c$. Because $M_s(T)$ exhibits thermal hysteresis (see Fig. 4 in Ref. 28) consistent with a first order transition, a fit of $M_s(T)$ to extract a power law dependence of the order parameter with temperature is not correct. For the film bulk, the shift of $T_c$ with -0.01% (compressive) strain is ~6 K. Thus, compressive strain promotes magnetism in LPCMO films to higher temperatures compared to the absence of applied stress. Similar increases of $T_c$ and the MIT with compressive strain is circumstantial evidence for an intimate relationship between the magnetic and metallic phases.

In summary, we found that bending stress producing 0.01% compressive strain increases the Curie temperature of the LPCMO film bulk by ~6 K—considerably larger than suggested by Millis *et al.*[18] Compressive strain also increases the metal-insulator-transition temperature by nearly the same amount (~4 K) as $T_c$. Thus, compressive strain favors the magnetic and metallic phases. Most importantly, the film retains significant magnetic order ~18 K above the metal-insulator-transition. We conclude that magnetic ordering is not caused by the metal-insulator-transition. When the magnetic ordering as measured by the ratio of saturation magnetization normalized to the 20 K value, i.e., $M_s(T)/M_s(20\ K)$, is less than ~0.6, the film's resistance measured over macroscopic dimensions changes from metallic to insulating. The value of ~0.6 is consistent with the site percolation threshold for two-dimensional spin lattices, thus, we suggest the onset of the metal-insulator transition maybe related to percolation of the magnetic phase if the magnetic phase is also metallic (a supposition supported by the similarity of increases of $T_c$ and the MIT with compressive strain).

This work was supported by the Office of Basic Energy Science (BES), U.S. Department of Energy (DOE), BES-DMS funded by the DOE's Office of BES, the National Science Foundation (DMR-0804452) (HJ and AB). Discussions with Profs. S.K. Sinha and Ivan K. Schuller and Drs. S. Guenon, G. Ramírez, T. Saerbeck are gratefully acknowledged. Los Alamos National Laboratory is operated by Los Alamos National Security LLC under DOE Contract DE-AC52-06NA25396.

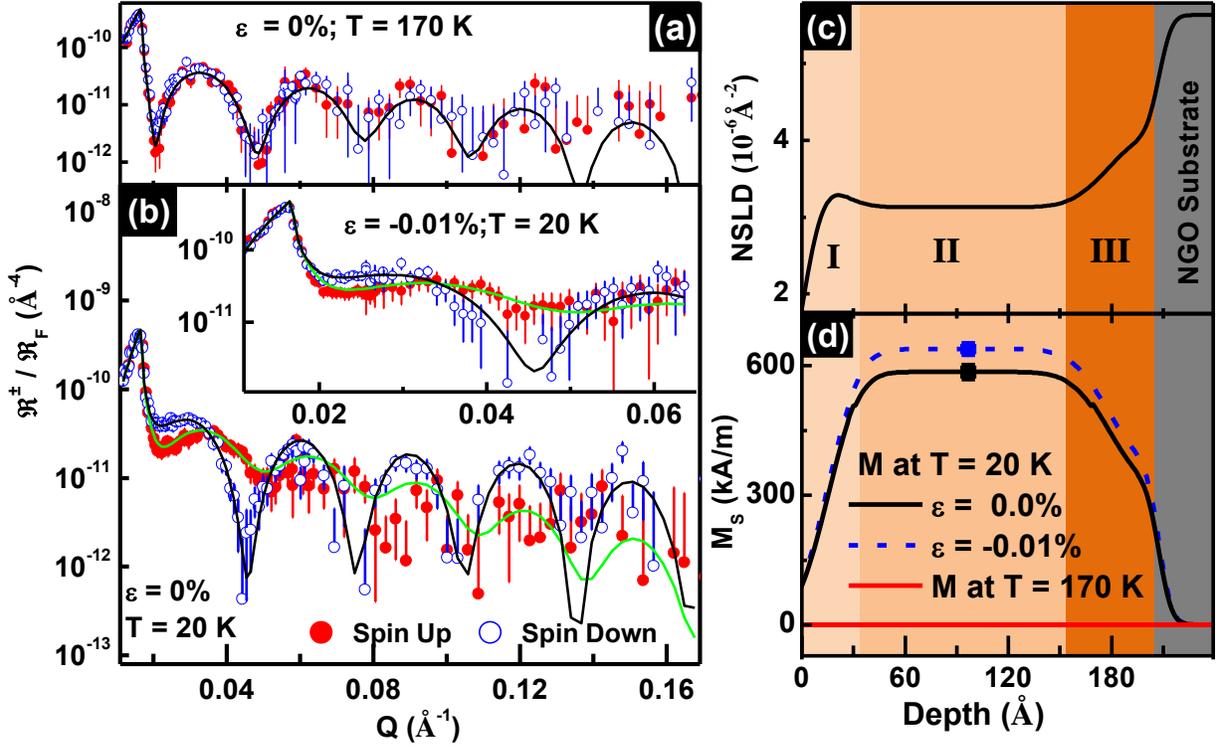

Fig. 1: PNR data normalized to the Fresnel reflectivity (symbols) from LPCMO film without applied stress at T = 170 K (a) and 20 K (b). Inset of (b) shows the PNR data from the LPCMO film at T = 20 K with applied compressive stress producing ε = -0.01%. (c) Nuclear scattering length density depth profile (NSLD) (d) and saturation magnetization depth profiles, which yield the solid curves (fit for PNR data, green/black) in (a) and (b).

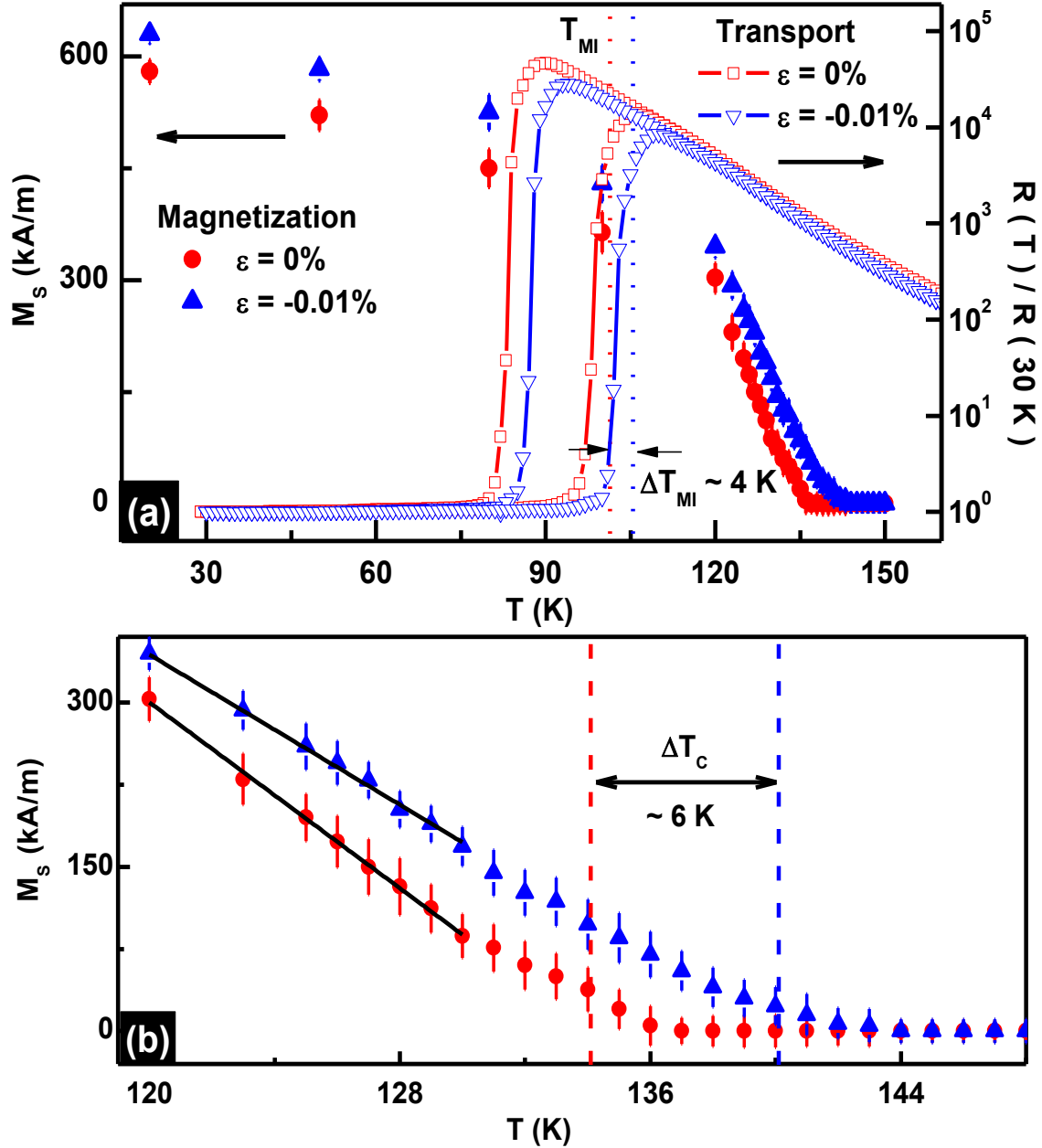

Fig. 2 (a) $M_s(T)$ obtained from the polarized neutron reflectivity of the LPCMO film without(●) and with (▲) applied bending stress while warming from low temperature. Superimposed is the resistance of the sample measured during the neutron experiment without (red) and with (blue) applied bending stress. The dotted lines correspond to the metal-insulator-transitions during warming for the two states of stress. (b) Shows the $M_s(T)$ close to the ordering temperature. Extrapolation of linear fits of $M_s(T)$ to $M_s = 0$ yields estimates for $T_c$.